\documentclass[%
 reprint,
 amsmath,amssymb,
 aps, prd, nofootinbib
]{revtex4-1}

\usepackage{graphicx}
\usepackage{dcolumn}
\usepackage{bm}
\usepackage{color}

\begin{document}

\title{Comment on ``Tidal Love numbers of neutron and self-bound quark stars''}

\author{J\'anos Tak\'atsy$^{1,2}$}
\email{takatsy.janos@wigner.hu}
\author{P\'eter Kov\'acs$^{1,2}$}%

\affiliation{
$^{1}$Institute for Particle and Nuclear Physics, Wigner Research Centre for Physics, 1121 Budapest, Konkoly-Thege Mikl\'os \'ut 29-33, Hungary
}%
\affiliation{%
$^{2}$Institute of Physics, E\"otv\"os University, 1117 Budapest, P\'azm\'any P\'eter stny. 1/A, Hungary
}%

\begin{abstract}
We comment on the paper of S. Postnikov et al. in Phys. Rev. D 82, 024016 (2010) and give a modified formula that needs to be taken into account when calculating the tidal Love number of neutron stars in case a first order phase-transition occurs at non-zero pressure. We show that the error made when using the original formula tends to zero as $p \rightarrow 0$ and we estimate the maximum relative error to be $\sim 5\%$ if the density discontinuity is at larger densities.
\end{abstract}

\maketitle

In Ref.~\cite{postnikov2010} the authors investigated the qualitative differences between the tidal Love numbers of self-bound quark stars and neutron stars. In Eq.~(14) they derived an expression for the extra term that should be subtracted from the logarithmic derivative $y(r)$ of the metric perturbation $H(r)$ in case there is a first-order phase transition in the equation of state (EoS). The authors applied this formula to quark stars where there is a core-crust phase transition at or below neutron-drip pressure. Since then multiple papers have included or applied this formula explicitly using EoSs with first-order phase transitions at non-negligible pressures ({\it e.g.} \cite{zhao2018,han2019}). However, when the pressure $p_d$ corresponding to the density discontinuity is non-negligible compared to the central energy density of the neutron star, Eq.~(14) of Ref.~\cite{postnikov2010} should be modified as shown below. In this comment we derive the correct formula and estimate the error made when using the other formula instead.

It needs to be added, that although Ref.~\cite{han2019} contains the uncorrected formula, the results presented in the paper were calculated using the correct relation, as it was reported by the authors and also verified by the authors of Ref.~\cite{postnikov2010}. This also applies to more recent publications including the same authors \cite{han2019b,chatziioannou2020}. Moreover, despite using the erroneous formula, the results of Ref.~\cite{zhao2018} are also mainly unaffected by this error, since they only provide approximate analytic fits for the ratios of tidal deformabilities of the two components in binary neutron stars. Thus, uncertainties of a few percent are inherently contained in these fits, which encompass the errors of individual tidal deformabilities. The corrected fits -- as it was claimed by the authors of Ref.~\cite{postnikov2010} -- are negligibly different from the reported fits in Ref.~\cite{zhao2018}. We also add that the correct formula appears in Ref.~\cite{zhang2020} as well.

The tidal $l=2$ tidal Love number can be expressed the following way:
\begin{align}
    k_2 &= \frac{8}{5} (1-2 \beta)^2 \beta^5 [2 \beta (y_R-1)-y_R+2]\nonumber\\
    &\times \{2 \beta [4 (y_R+1) \beta^4+(6 y_R-4) \beta^3+(26-22 y_R) \beta^2\nonumber\\
    &+3 (5 y_R-8) \beta-3 y_R+6]+3 (1-2 \beta)^2\nonumber\\
    &\times[2 \beta (y_R-1)-y_R+2]\ln \left(1-2\beta\right)\}^{-1} ,
\label{eq:k2}
\end{align}
where $\beta=M/R$ is the compactness parameter of the neutron star and $y_R=y(R)=[rH'(r)/H(r)]_{r=R}$ with $H(r)$ being a function related to the quadrupole metric perturbation (see {\it e.g.} \cite{damour2009}). $y_R$ is obtained by solving the following first-order differential equation:
\begin{align}
    ry'(r)&+y(r)^2+r^2 Q(r) \nonumber\\
    &+ y(r)e^{\lambda(r)}\left[1+4\pi r^2(p(r)-\varepsilon(r))\right] = 0 ,
\label{eq:y}
\end{align}
where $\varepsilon$ and $p$ are the energy density and pressure, respectively, and
\begin{align}
    Q(r)=4\pi e^{\lambda(r)}\left(5\varepsilon(r)+9p(r)+\frac{\varepsilon(r)+p(r)}{c_s^2(r)}\right) \nonumber\\
    -6\frac{e^{\lambda(r)}}{r^2}-(\nu'(r))^2 .
\label{eq:Q}
\end{align}
Here $c_s^2=\mathrm{d}p/\mathrm{d}\varepsilon$ is the sound speed squared, while $e^{\lambda(r)}$, $\nu(r)$ metric functions are given by
\begin{align}
e^{\lambda(r)} &= \left[1-\frac{2m(r)}{r}\right]^{-1} \label{eq:tov_e} , \\ 
\nu'(r) &= \dfrac{2[m(r)+4\pi r^3 p(r)]}{r^2 - 2 m(r) r} \label{eq:tov_nu} ,
\end{align}
with the line element for the unperturbed star defined as
\begin{equation}
    \mathrm{d}s^2 =  e^{\nu(r)}\mathrm{d}t^2 - e^{\lambda(r)}\mathrm{d}r^2 - r^2(\mathrm{d}\vartheta^2 + \sin^2\vartheta \, \mathrm{d}\varphi^2),
\end{equation}
and where $m(r)$ and $p(r)$ are calculated through the Tolman-Oppenheimer-Volkoff equations \cite{tolman1939,oppenheimer1939}:
\begin{align}
m'(r) &= 4\pi r^2 \varepsilon(r) , \label{eq:tov_m} \\ 
p'(r) &= - [\varepsilon(r)+p(r)]\dfrac{m(r)+4\pi r^3 p(r)}{r^2 - 2 m(r) r} .\label{eq:tov_p}
\end{align}

In case there is a first-order phase transition in the EoS, there is a jump of $\Delta\varepsilon$ in the energy density at constant pressure, hence $c_s^2=0$ in that region and the term in Eq.~(\ref{eq:Q}) containing $1/c_s^2$ diverges. Expressing $1/c_s^2$ in the vicinity of the density discontinuity:
\begin{equation}
    \frac{1}{c_s^2} = \frac{\mathrm{d}\varepsilon}{\mathrm{d}p}\bigg|_{p\neq p_d} + \delta(p-p_d) \Delta \varepsilon .
\label{eq:cs2}
\end{equation}
Changing the delta-function to a function in the radial position $r$, inserting Eq. (\ref{eq:cs2}) into Eq. (\ref{eq:y}) and integrating over an infinitesimal distance around $r_d$ one obtains:
\begin{equation}
    y(r_d^+) - y(r_d^-) = -4\pi r_d e^{\lambda(r_d)} [\varepsilon(r_d)+p(r_d)] \frac{\Delta \varepsilon}{|p'(r_d)|} .
\end{equation}
Using Eq.~(\ref{eq:tov_p}) we get:
\begin{align}
    y(r_d^+) - y(r_d^-) &= -\frac{4\pi r_d^3 \Delta \varepsilon}{m(r_d)+4\pi r_d^3 p(r_d)}\nonumber\\
    &= -\frac{\Delta \varepsilon}{\tilde{\varepsilon}/3+p(r_d)} ,
\label{eq:ydisc}
\end{align}
where $\tilde{\varepsilon}=m(r_d)/(4\pi r_d^3/3)$ is the average energy density of the inner ($r<r_d$) region. Eq.~(\ref{eq:ydisc}) shows that there is an extra $p(r_d)$ term in the denominator as compared to Eq.~(14) of Ref.~\cite{postnikov2010}. We see that if the phase transition is at very low densities compared to the central energy density then $p(r_d)/\tilde{\varepsilon}\rightarrow0$ \footnote{It is worth to note here that although $\tilde{\varepsilon}$ -- the average energy density of the inner core -- is not equal to the central energy density $\varepsilon_c$, it falls to the same order of magnitude ($\tilde{\varepsilon}/\varepsilon_c \gtrsim 0.25 - 0.5$ for $M>0.5$~$M_\odot$).} and we get back the formula in Ref.~\cite{postnikov2010}.

\begin{figure}[t]
\includegraphics[width=0.48\textwidth]{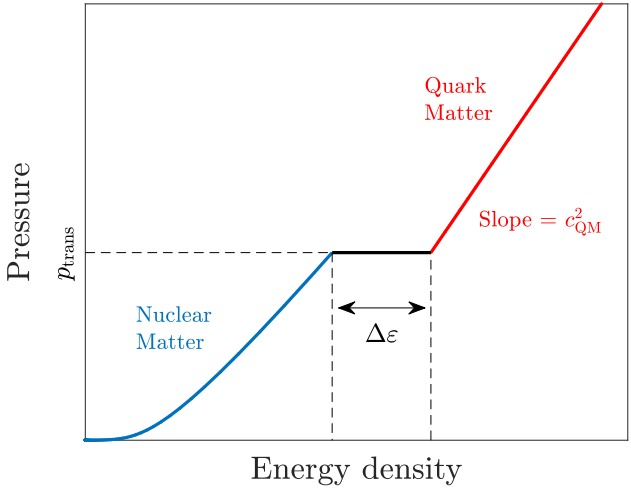}
\caption{\label{fig:css}Illustration of the EoS in the constant-sound-speed construction \cite{alford2013,han2019}. At $p_\mathrm{trans}$ a quark matter part with a constant sound speed of $c_\mathrm{QM}$ is attached to the nuclear matter EoS after an energy density jump of $\Delta\varepsilon$.}
\end{figure}

We investigated the difference caused by applying the two different formulas using a constant-sound-speed construction (see Fig.~\ref{fig:css}) \cite{alford2013,han2019}:
\begin{equation}
\varepsilon(p)=
    \bigg\{\begin{array}{lr}
    \varepsilon_\mathrm{NM}(p) &p<p_\mathrm{trans}\\
    \varepsilon_\mathrm{NM}(p_\mathrm{trans}) + \Delta \varepsilon + c_\mathrm{QM}^{-2}(p-p_\mathrm{trans}) \quad &p>p_\mathrm{trans}
    \end{array},
\end{equation}
where we fixed $c_\mathrm{QM}^2 = 1$ as in Ref.~\cite{han2019}, while varying the values of $p_\mathrm{trans}$ (through $n_\mathrm{trans}\equiv n_\mathrm{NM}(p_\mathrm{trans})$) and $\Delta \varepsilon$. For the nuclear matter (NM) part we chose the Steiner--Fischer--Hempel (SFHo) EoS \cite{steiner2013} and the Hempel–-Schaffner-Bielich EoS with density-dependent relativistic mean-field interactions (DD2) \cite{typel2010,hempel2010} as two representative EoSs. We varied the baryon number density at the phase transition $n_\mathrm{trans}$ between $n_0$ and $3.5 n_0$ with $n_0=0.16$~fm$^{-3}$ being the nuclear saturation density. The strength of the phase transition $\Delta\varepsilon/\varepsilon_\mathrm{trans}$ was varied between $0$ and $3$, where $\varepsilon_\mathrm{trans}\equiv\varepsilon_\mathrm{NM}(p_\mathrm{trans})$.

\begin{figure}[t]
\includegraphics[width=0.48\textwidth]{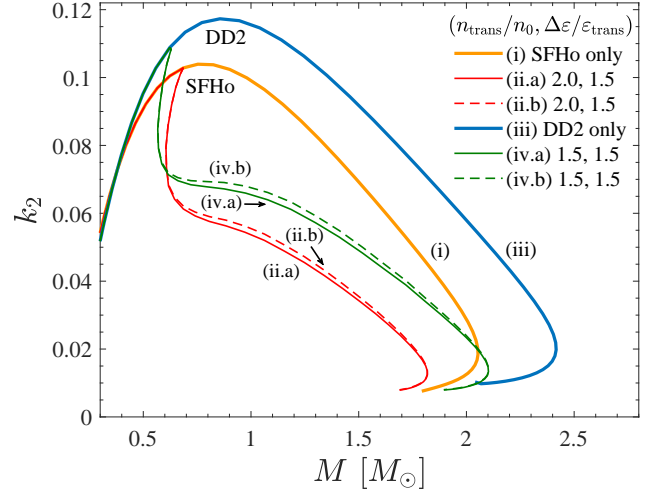}
\caption{\label{fig:k2ex}Tidal Love number--neutron star mass relations for the SFHo (orange line) and DD2 (blue line) EoSs, as well as for EoSs obtained from the constant-sound-speed construction. The different number pairs denote different values of $n_\mathrm{trans}/n_0$ and $\Delta\varepsilon/\varepsilon_\mathrm{trans}$, respectively. The tidal Love numbers calculated using Eq.~(\ref{eq:ydisc}) (solid lines) are reduced by a few percent compared to the ones calculated using Eq.~(14) of Ref.~\cite{postnikov2010} (dashed lines).}
\end{figure}

\begin{figure*}[!htb]
\includegraphics[width=0.48\textwidth]{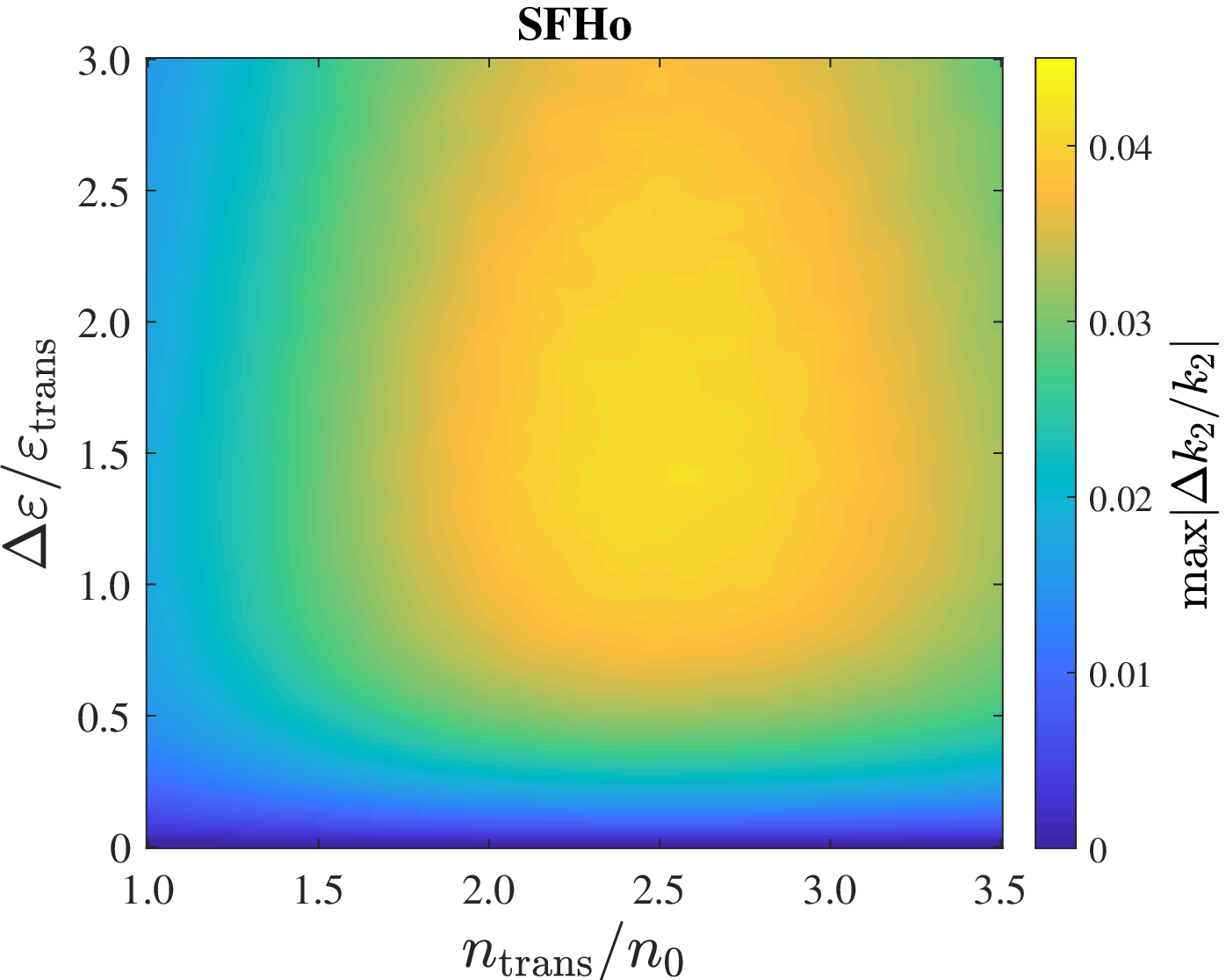}
\hfill
\includegraphics[width=0.48\textwidth]{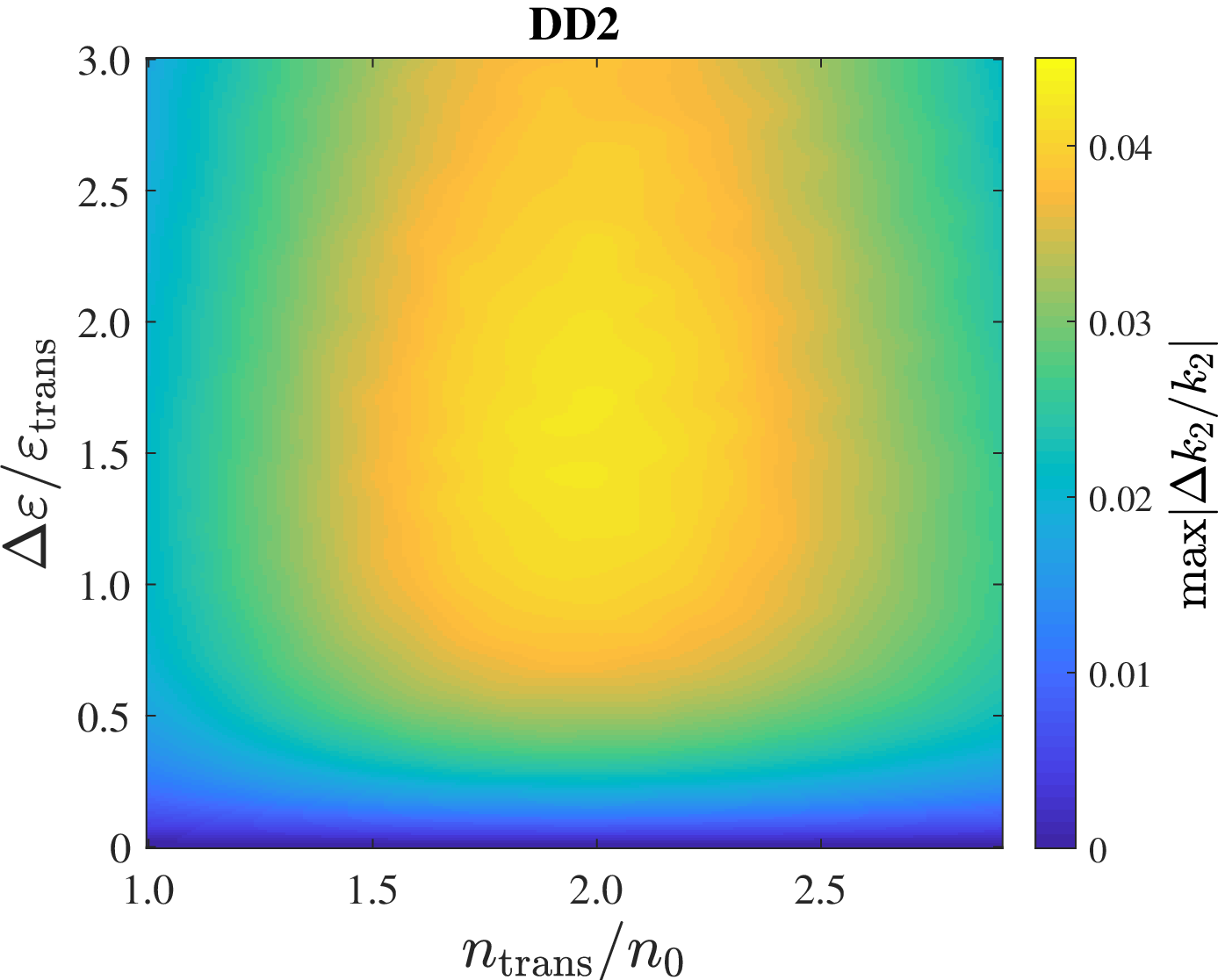}
\caption{\label{fig:k2err}The maximum relative difference between the tidal Love numbers calculated using the two different equations with the SFHo (left panel) and DD2 (right panel) nuclear matter EoSs in a constant-sound-speed construction. The maximum value is $\sim4.2\%$ for the SFHo and $\sim4.3\%$ for the DD2 EoS. Constant-sound-speed constructions with $n_\mathrm{trans}/n_0>3.0$ for the DD2 EoS mostly contain mass-radius relations with no stable hybrid star branches and hence were omitted from the figure.}
\end{figure*}

In Fig.~\ref{fig:k2ex} we show some examples of tidal Love number--neutron star mass relations. For EoSs with first-order phase transitions, the Love numbers are reduced when using Eq.~(\ref{eq:ydisc}) (red and green solid lines) compared to using the formula in Ref.~\cite{postnikov2010} (red and green dashed lines). The maximum relative difference in the tidal Love number as a function of the two parameters defining our constant-sound-speed EoSs is shown in Fig.~\ref{fig:k2err}. We see that the maximum relative difference reaches its maximum at $n_\mathrm{trans}/n_0\approx2.5$ and $\Delta\varepsilon/\varepsilon_\mathrm{trans}\approx1.5$ for the SFHo EoS, and at $n_\mathrm{trans}/n_0\approx2.0$ and $\Delta\varepsilon/\varepsilon_\mathrm{trans}\approx1.5$ for the DD2 EoS, however, it does not exceed $5\%$ for the whole parameter range. The relative difference also diminishes as we go to lower densities, as it is expected. \\

J. T. and P. K. acknowledge support by the National Research, Development and Innovation (NRDI) fund of Hungary, financed under the FK\_19 funding scheme, Project No. FK 131982. P. K. also acknowledges support by the János Bolyai Research Scholarship of the Hungarian Academy of Sciences.


\end{document}